\documentclass{ieeetj}
\usepackage{cite}
\usepackage{amsmath,amssymb,amsfonts}
\usepackage{algorithmic}
\usepackage{graphicx,color}
\usepackage{textcomp}
\usepackage{xcolor}
\usepackage{hyperref}
\hypersetup{hidelinks=true}
\usepackage{algorithm,algorithmic}
\def\BibTeX{{\rm B\kern-.05em{\sc i\kern-.025em b}\kern-.08em
    T\kern-.1667em\lower.7ex\hbox{E}\kern-.125emX}}
\AtBeginDocument{\definecolor{tmlcncolor}{cmyk}{0.93,0.59,0.15,0.02}\definecolor{NavyBlue}{RGB}{0,86,125}}

\def\OJlogo{\vspace{-4pt}$<$Society logo(s) and publication title will appear here.$>$}
\def\seclogo{\vspace{10pt}$<$Society logo(s) and publication title will appear here.$>$}

\def\authorrefmark#1{\ensuremath{^{\textbf{#1}}}}

\begin{document}
\receiveddate{XX Month, XXXX}
\reviseddate{XX Month, XXXX}
\accepteddate{XX Month, XXXX}
\publisheddate{XX Month, XXXX}
\currentdate{XX Month, XXXX}
\doiinfo{XXXX.2022.1234567}
\markboth{}{Caulfield M. {et al.}}

\title{Improving the Elevational Focusing of Fast Orthogonal Row-Column Electronic Scanning (FORCES) Ultrasound Imaging using Retrospective Transmit Beamforming (RTB)}
\author{Michael Caulfield\authorrefmark{1}, Randy Palamar\authorrefmark{1}, Graduate Student Member, IEEE, Darren Dahunsi\authorrefmark{1}, Graduate Student Member, IEEE, Mohammad Rahim Sobhani\authorrefmark{1}, Member, IEEE, Negar Majidi\authorrefmark{1}, Member, IEEE, and Roger Zemp\authorrefmark{1}, Member, IEEE}

\affil{Faculty of Engineering, University of Alberta, AB T6G 2R3, Canada}

\corresp{Corresponding authors: Michael Caulfield (email: mcaulfie@ualberta.ca), Roger Zemp (email: rzemp@ualberta.ca).}
\authornote{Funding for this work was provided by the National Institutes of Health (1R21HL161626-01 and EITTSCA R21EYO33078), Alberta Innovates(AICE 202102269, CASBE 212200391 and LevMax 232403439), MITACS (IT46044 and IT41795), CliniSonix Inc., NSERC (2025-05274),the Alberta Cancer Foundation and the Mary Johnston Family Melanoma Grant (ACF JFMRP 27587), the Government of Alberta CancerResearch for Screening and Prevention Fund (CRSPPF 017061), an Innovation Catalyst Grant to MRS, and INOVAIT (2023-6359). We aregrateful to the nanoFAB staff at the University of Alberta for facilitating array fabrication.}

\begin{abstract}
Recent developments in Row Column Arrays (RCAs) have presented promising options for volumetric imaging without the need for the excessive channel counts of fully wired 2D-arrays. Bias programmable RCAs, also known as Top Orthogonal to Bottom Electrode (TOBE) Arrays, show further promise in that imaging schemes, such as Fast Orthogonal Row-Column Electronic Scanning (FORCES) allow for full transmit and receive focusing everywhere in the image plane. However, due to its fixed elevational focus and large transmit aperture, FORCES experiences poor elevational focusing away from the focal point. In this study we present a modification to the FORCES imaging scheme by applying Retrospective Transmit Beamforming (RTB) in the elevational direction to allow for elevational transmit focusing everywhere in the imaging plane. We evaluate FORCES and uFORCES methods, with and without RTB applied, when imaging both a cyst and wire phantom. With experiment we show improved elevational focusing capabilities away from the focal point when RTB is applied to both FORCES and uFORCES. At the focal point, performance with RTB remains comparable or improved relative to standard FORCES. This is quantified by the measurement of Full Width Half Max when imaging the wire phantom, and by the generalized Contrast to Noise Ratio when imaging the tubular cyst phantom. We also demonstrate the volumetric imaging capabilities of FORCES RTB with the wire phantom.
\end{abstract}


\maketitle

\section{INTRODUCTION}
\label{sec:introduction}
\IEEEPARstart{D}{espite} improvements in 3-D or fully electronically scannable ultrasound imaging technology, limitations remain. Most fully wired matrix probes are limited to 32$\times$32 element arrays or require complex micro-beamformers. This leads to limitations in Field of View (FOV), resolution, and signal to noise ratio (SNR) \cite{32x32Probe}, \cite{3dimaging}. Row-Column Arrays (RCAs) provide a promising option for 3-D imaging without the need for excessive channel counts. However, standard piezoelectric RCAs cannot achieve full transmit and receive focusing and cannot perform imaging outside the shadow of the aperture without the use of an acoustic lens \cite{LensedRCA}, \cite{FORCESvsRCA_PrePrint}.

Bias sensitive RCAs or Top Orthogonal to Bottom Electrode (TOBE) arrays are similar to standard RCAs. However, they are manufactured with electrostrictive relaxors, such as PMN-PT that exhibit a piezoelectric effect roughly proportional to the polarity and magnitude of the DC bias voltage applied across the relaxor \cite{Electrostrictive}. When combined with programmable DC biasing electronics, this property allows for more complex imaging schemes with improved resolution and the ability to image outside the shadow of the aperture \cite{HVBiasElectronics}. 

\begin{figure}[!t]
\centerline{\includegraphics[width=\columnwidth]{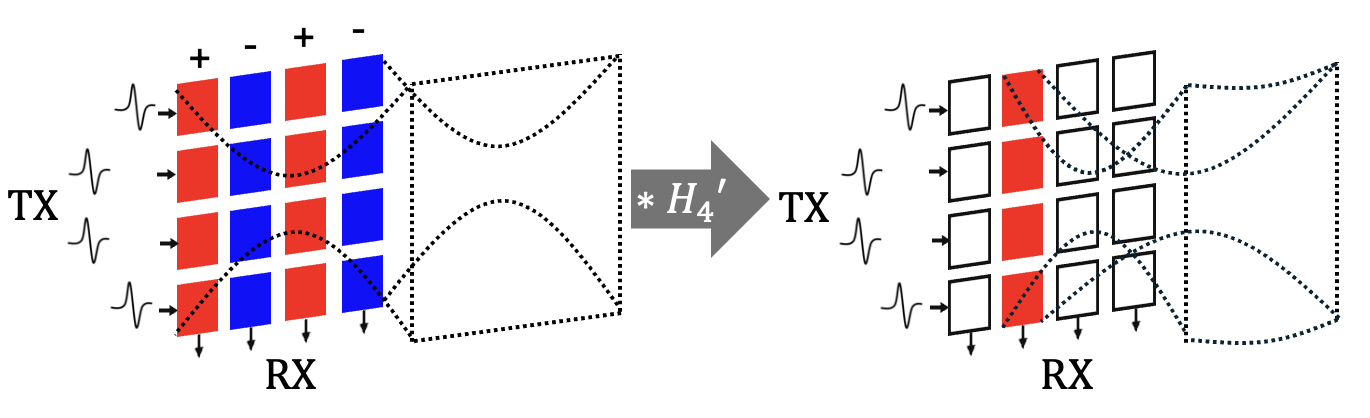}}
\caption{Second transmit before (left) and after (right) decoding of a FORCES scheme for a 4$\times$4 TOBE array. Red, blue, and white represent an element biased positive, negative, and grounded respectively.}
\label{fig1}
\end{figure}

Fast Orthogonal Row-Column Electronic Scanning (FORCES) is one such imaging scheme, which can achieve transmit and receive focusing everywhere in the imaging plane \cite{FORCES},\cite{uFORCES}. Unlike conventional RCAs, FORCES can also perform imaging outside the shadow of the aperture \cite{FORCESvsRCA_PrePrint}. FORCES achieves this by emitting an elevationally focused transmit along the rows of a TOBE Array while applying a biasing pattern along the columns with the same polarity as the columns of a Hadamard matrix. The data is then decoded by multiplication of the transposed Hadamard matrix. This process gives an aperture-encoded version of Synthetic Aperture (SA) imaging with an electronically steerable elevational focus. SA imaging allows for full transmit and receive focusing everywhere in the imaging plane \cite{JENSEN2006e5}, \cite{AppSAImaging}.  Figure~\ref{fig1} shows the transmit patterns (actual and effective after decoding) for the second transmit of a 4$\times$4 element TOBE array.

Despite the impressive quality of FORCES images, the large transmit aperture and fixed elevational focus make for significant out of plane artifacts. Similarly, the fixed elevational focus means each FORCES acquisition produces only a B-Scan image. For a 128$\times$128 element array, 128 transmits are required to produce one FORCES image, this makes even thin slab volume acquisitions impractical. Other methods such as uFORCES allow for a faster acquisition rate, allowing for an image to be created with fewer transmits, though the quality is reduced when compared to FORCES \cite{uFORCES}.

The use of a Virtual Source (VS) for ultrasound imaging has been utilized for some SA imaging methods and RCA imaging schemes \cite{JENSEN2006e5}, \cite{SA_2}, \cite{VLS_apo}. The use of a virtual source has also been suggested for improved elevational focusing when creating a volume from a series of phased array imaging scan planes \cite{VirtualSource}.

In this paper we present FORCES Retrospective Transmit Beamforming (RTB), a modification to the FORCES imaging scheme which aims to improve the elevational focusing by implementing RTB in the elevational direction. This gives simultaneous SA imaging in the lateral direction and VS imaging in the elevational direction. We show an experimental improvement in elevational focusing without a significant reduction in image quality when compared to standard FORCES imaging methods. 

\section{METHODS}
\label{sec:methods}
\subsection{Imaging System}
\begin{figure}[!t]
\centerline{\includegraphics[width=\columnwidth]{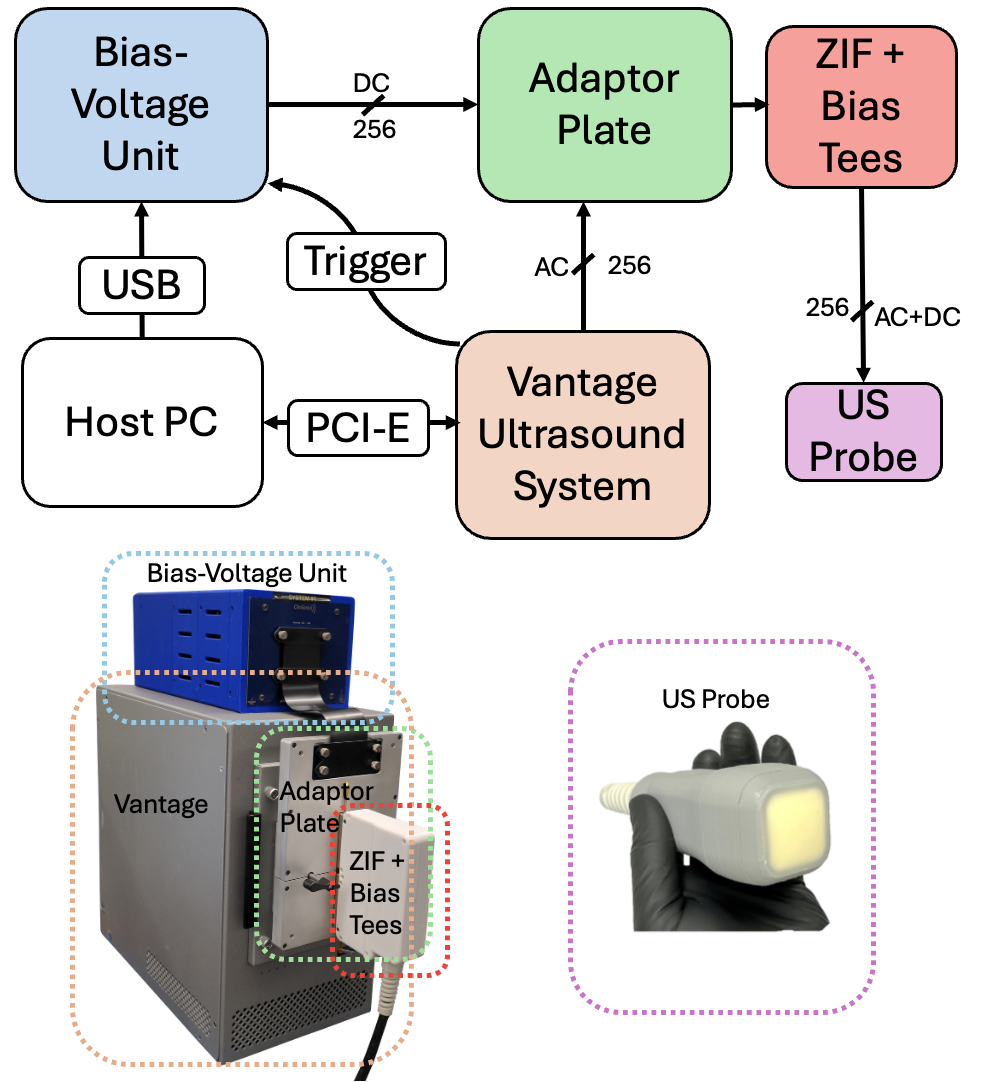}}
\caption{TOBE Array imaging setup. Top shows a simplified block diagram of the system and the bottom shows the experimental setup}
\label{fig_setup}
\end{figure}

A handheld 128$\times$128 element CliniSonix (Edmonton, Canada) TOBE array with $\lambda$-pitch and a center frequency of 6.2 MHz was used to perform the imaging. The imaging system also included a Verasonics (Kirkland, USA) Vantage System, a CliniSonix adapter plate, and a CliniSonix bias-voltage unit. The Vantage System was used to provide elevationally focused AC transmit signals, the bias-voltage unit was used to supply the DC bias voltages, and the adapter plate was used to connect the array, Vantage System, and bias-voltage unit. Figure \ref{fig_setup} shows the setup used for the experiment.

\subsection{Data Acquisition}
The Sun Nuclear (Melbourne, USA) ATS-539 quality assurance phantom was imaged with the CliniSonix Array at 4.3 MHz for cyst and wire targets. Because the focus of this study is on elevational focus capabilities, the phantom was oriented so that the tubular cysts and wires are parallel to the imaging plane. FORCES images were acquired using the imaging system described above. A walking FORCES scheme was used, in which the elevational focal zone was shifted in increments of 2$\times$pitch across the array. This gave 64 FORCES image slices, each spaced 500 $\mu$m apart. The data were decoded using the appropriate matrix multiplication. uFORCES, FORCES RTB, and uFORCES RTB were emulated by selecting the correct decoded data from FORCES. This was done to ensure that there were no differences in the position of the transducer or random noise when comparing the methods. 

For a 128$\times$128 element array, FORCES requires 128 transmits and yields 128 usable SA transmits after decoding. uFORCES requires 16 transmits and and yields 15 usable SA transmits after decoding. As such, 15 transmits from the decoded FORCES data set were used to beamform the image. For this study, data from 16 imaging planes were used to form each image of the RTB imaging schemes. For uFORCES RTB 15 transmits from 16 different imaging planes were used, and for FORCES RTB 128 transmits from 16 different imaging planes were used. 16 imaging planes were chosen as improvements were marginal when using more. A summary of the the 4 imaging schemes can be found in Table \ref{tab:imaging_schemes}.

\begin{table}[t]
\caption{Comparison of imaging schemes for a 128$\times$128 element TOBE Array. A pulse repetition frequency of 4kHz is assumed when calculating frame rate.}

\centering
\resizebox{\columnwidth}{!}{
\begin{tabular}{lcccc}
\hline
\textbf{Imaging Scheme} & 
\begin{tabular}[c]{@{}c@{}}\textbf{Transmits}\\\textbf{Before}\\\textbf{Decoding}\end{tabular} &
\begin{tabular}[c]{@{}c@{}}\textbf{Effective SA}\\\textbf{Transmits}\\\textbf{After Decoding}\end{tabular} &
\begin{tabular}[c]{@{}c@{}}\textbf{\# of Imaging}\\\textbf{Planes}\\\textbf{Beamformed}\end{tabular} &
\begin{tabular}[c]{@{}c@{}}\textbf{Frame}\\\textbf{Rate (fps)}\end{tabular} \\
\hline
FORCES       & 128  & 128  & 1  & 31  \\
uFORCES 16   & 16   & 15   & 1  & 250 \\
uFORCES RTB  & 256  & 240  & 16 & 16  \\
FORCES RTB   & 2048 & 2048 & 16 & 2   \\
\hline
\end{tabular}
}
\label{tab:imaging_schemes}
\end{table}

\subsection{Beamforming}

\begin{figure}[!t]
\centerline{\includegraphics[width=\columnwidth]{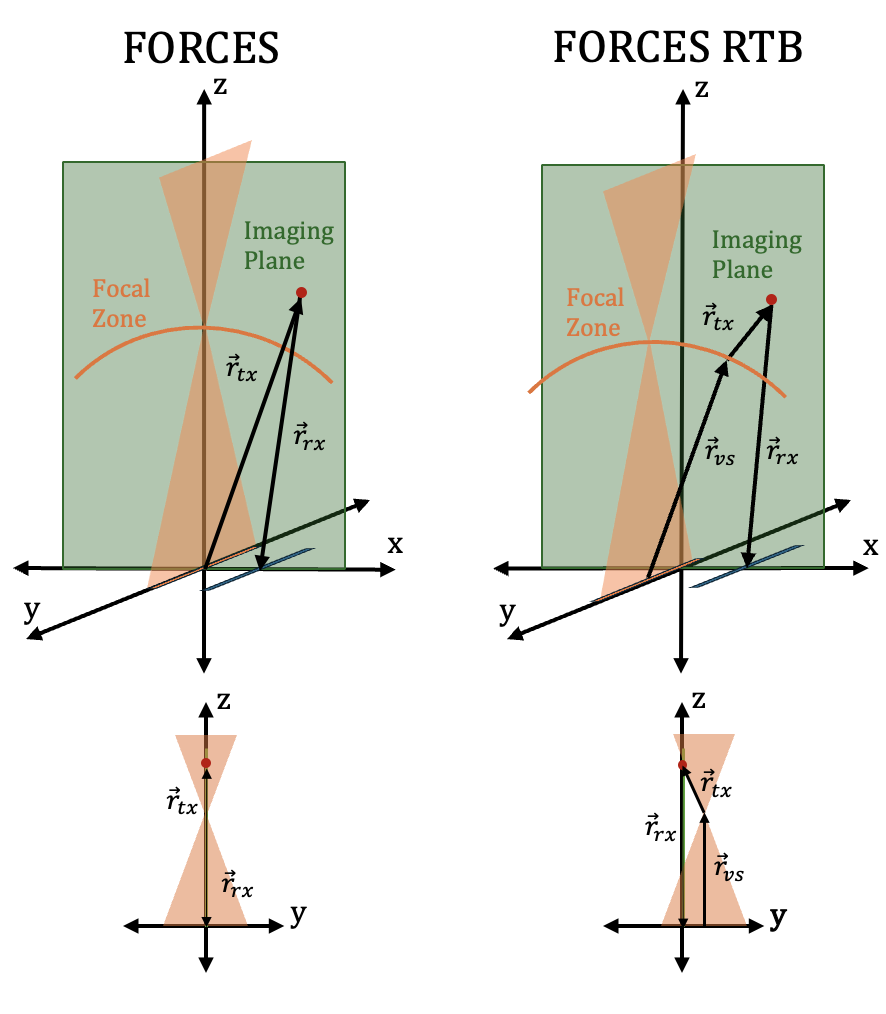}}
\caption{Left shows the standard 2D FORCES imaging geometry where an effective single column transmit array is assumed (after Hadamard aperture decoding). Here the y direction is the elevational direction, x is the azimuthal direction, and z is depth. An elevational focus is achieved with transmit delays applied to rows. Hence the focal zone for a single column transmitter is approximated by an arc in the azimuthal slice. FORCES RTB enables elvationally scanned FORCES imaging to achieve 3D imaging with retrospective transmit focusing. This treats the focal zone as an arc shaped virtual source. In the FORCES RTB imaging geometry, the emission plane is not necessarily in the imaging plane (as is the case with FORCES). After a series of walking emisions in the elevational direction, a retrospective beamforming algorithm is used to form a 3D image.}
\label{fig2}
\end{figure}

Four different imaging and beamforming schemes were compared, FORCES, uFORCES, FORCES RTB, and uFORCES RTB. In short, the RTB methods differ from the standard FORCES methods by treating the elevational focal arc as a virtual source and moving it in and out of the imaging plane. Thus allowing for elevational transmit focusing everywhere in the imaging plane. The transmit path length calculation and transmit sequence are adjusted to achieve this. For standard FORCES, the path length is calculated as follows, where $\vec{r}_{rx}$ is the receive path length, $\vec{r}_{tx}$ is the transmit path length, $\vec{x}_{px}$ is the pixel location, $\vec{x}_{tx}$ is the center of the transmitting element, and the $\vec{x}_{rx}$ is the center of the receiving element. Further, for standard FORCES, the elevational focus is always on the imaging plane.
\begin{equation}
\label{eq:eq1}
d_{FORCES} = |\vec{r}_{rx}| + |\vec{r}_{tx}| = |\vec{x}_{rx} - \vec{x}_{px}| + |\vec{x}_{px} - \vec{x}_{tx}|
\end{equation}
For FORCES RTB, the equation is adjusted such that it is assumed the transmit path must pass through the virtual source. Following decoding, each transmit event acts as if an elevationally focused transmit occurred on one long element. As such, for the purposes of beamforming, it is assumed there is no directivity or focusing in the lateral direction. This gives an arc with a radius of the focal distance as the virtual source.  
With these assumptions, the total path length can be calculated as follows, where $\vec{r}_{vs}$ is the path from the transmitting element to the virtual source, $\vec{r}_{tx}$ is the path from the virtual source to the beamformed pixel, and $f_{dist}$ is the focal distance. 
\begin{equation}
\label{eq:eq2}
d_{FORCES_{RTB}} = |\vec{r}_{rx}| + |\vec{r}_{vs}| \pm |\vec{r}_{tx}|
\end{equation}

Because the virtual source is an arc, $|\vec{r}_{vs}|$ is always equal to $f_{dist}$. To find a generalized equation, $\vec{r}_{tx}$ is split into $\vec{r}_{tx_{x,z}}$ and $\vec{r}_{tx_{y}}$. This gives

\begin{equation}
\label{eq:eq3}
d_{FORCES_{RTB}} = |\vec{x}_{px} - \vec{x}_{rx}| + f_{dist} \pm |\vec{r}_{tx_{x,z}}+\vec{r}_{tx_y}|
\end{equation}
Where
\begin{equation}
\label{eq:eq4}
|\vec{r}_{tx_{x,z}}| = \sqrt{(x_{tx_x}-x_{px_x})^2 + (x_{tx_z}-x_{px_z})^2} - f_{dist}
\end{equation}
And
\begin{equation}
\label{eq:eq5}
|\vec{r}_{tx_{y}}| = |x_{tx_y} - x_{px_y}|
\end{equation}
Finally, substituting (\ref{eq:eq4}) and (\ref{eq:eq5}) into (\ref{eq:eq3}) we get
\begin{equation}
\label{eq:eq6}
\begin{aligned}
&d_{\mathrm{FORCES\_RTB}}
= \lVert \vec{x}_{px} - \vec{x}_{rx} \rVert + f_{dist} \\
&\qquad \pm \Biggl(
   \bigl(
      \sqrt{(x_{tx_x}-x_{px_x})^{2} + (x_{tx_z}-x_{px_z})^{2}}
      - f_{dist}
   \bigr)^{2} \\
&\qquad 
   + (x_{tx_y} - x_{px_y})^{2}
   \Biggr)^{\tfrac{1}{2}}
\end{aligned}
\end{equation}
The ± is determined by whether the beamformed pixel is above or below the focal zone. Figure \ref{fig2} demonstrates the difference between the two beamforming methods. A constant F-number of 1 was used for apodization on both transmit and receive.

\section{RESULTS}
\label{sec:results}
\subsection{Resolution Targets}
\begin{figure}[!t]
\centerline{\includegraphics[width=\columnwidth]{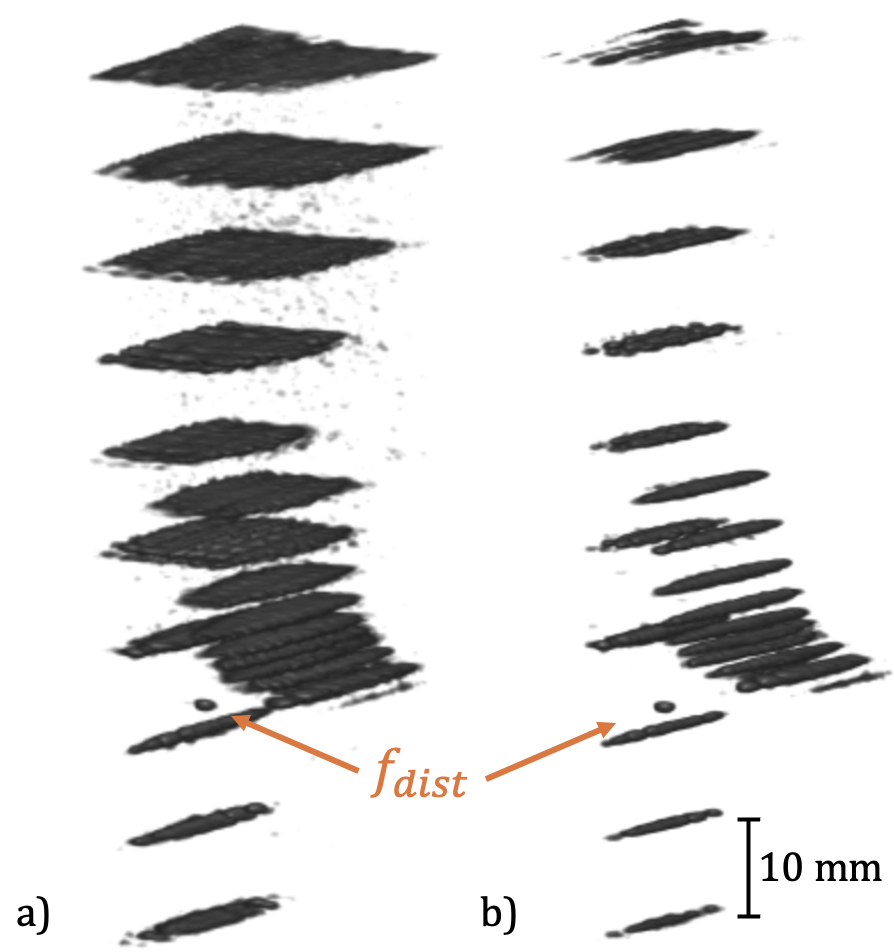}}
\caption{Wire phantom volume reconstruction using a) Walking FORCES and b) uFORCES RTB. The focal zone is indicated by the orange arrows.}
\label{fig3}
\end{figure}

\begin{figure}[!t]
\centerline{\includegraphics[width=\columnwidth]{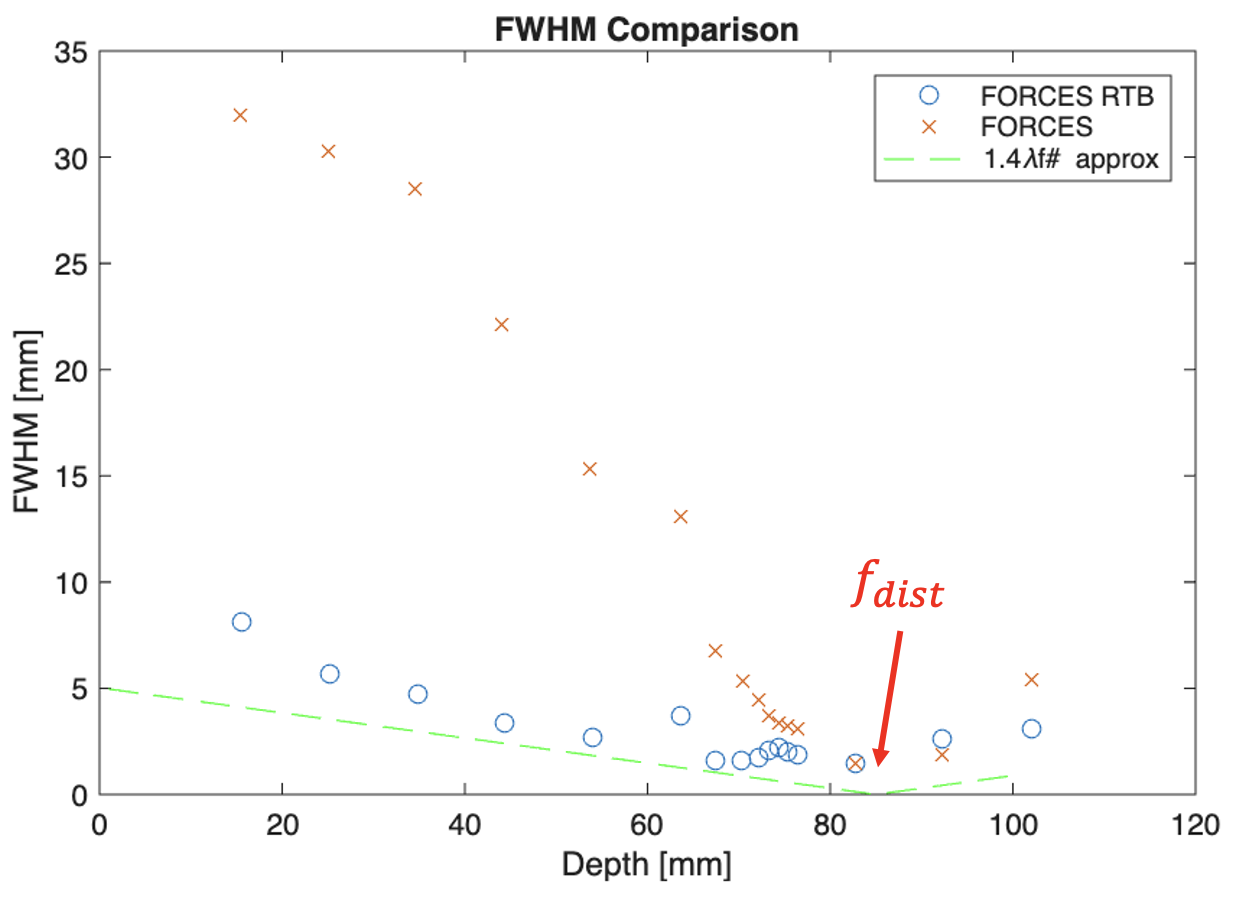}}
\caption{Elevational FWHM measured on a wire phantom, comparing Walking FORCES and uFORCES RTB.}
\label{fig4}
\end{figure}

Walking FORCES and uFORCES RTB were compared when capturing a 64 slice volume of a wire phantom. Each wire was run parallel to the image slices. Figure \ref{fig3} shows a volume reconstruction from the two methods. Notably, the uFORCES RTB method requires 1/8th the transmits as the FORCES method to construct the same volume. Figure \ref{fig4} shows the Full Width Half Max (FWHM) comparison between the two methods. Both methods achieve similar resolution near the focal point, FWHM of 1.5 mm for both at a distance of 2.2 mm from the focal point. However, at further distances uFORCES RTB outperforms FORCES with a FWHM of 4.7 mm compared to 28.5 mm respectively at a distance of 50 mm from the focal zone. The course resolution of FORCES is simply because we are at a point far from the elevational transmit focus. While an improvement is seen with RTB, the elevational resolution is still not perfect, but as can be seen in Figure \ref{fig4}, the FWHM for uFORCES RTB follows the expected 1.4$\lambda$$f_\#$ approximation for ultrasound resolution \cite{FWHM}.

\subsection{Contrast Targets}

\begin{figure}[!t]
\centerline{\includegraphics[width=\columnwidth]{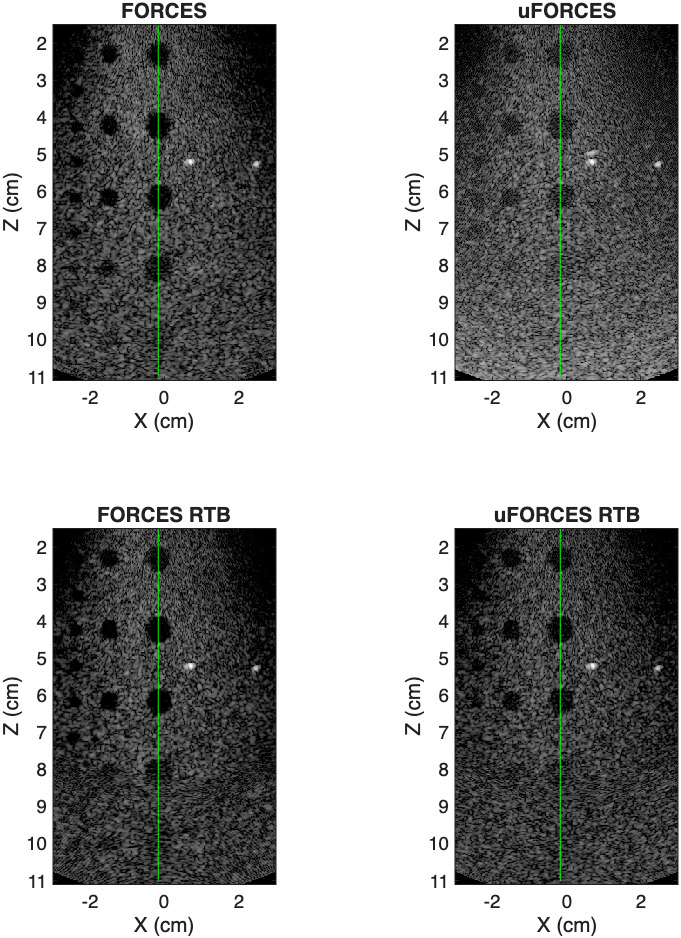}}
\caption{Perpendicular View of the cysts analized in this study. The imaging plane utilized is shown by the green line. The elevational focal zone is at 85 mm}
\label{figcystref}
\end{figure}

\begin{figure}[!t]
\includegraphics[width=\columnwidth]{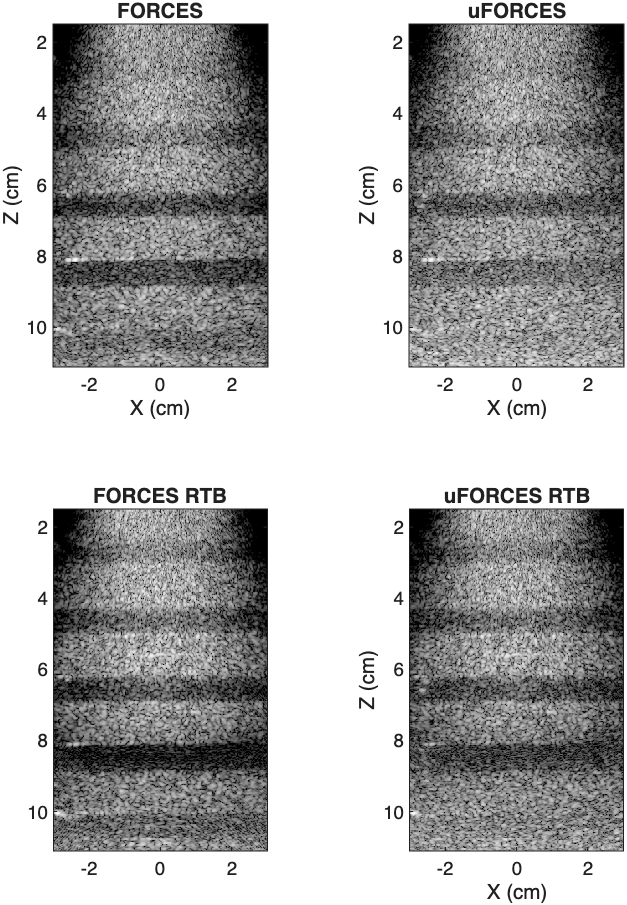}
\centering
\caption{Tubular cysts imaged at 4.3 MHz with FORCES, uFORCES, uFORCES RTB, and FORCES RTB.}
\label{fig5}
\end{figure}
\begin{figure}[!t]
\centerline{\includegraphics[width=\columnwidth]{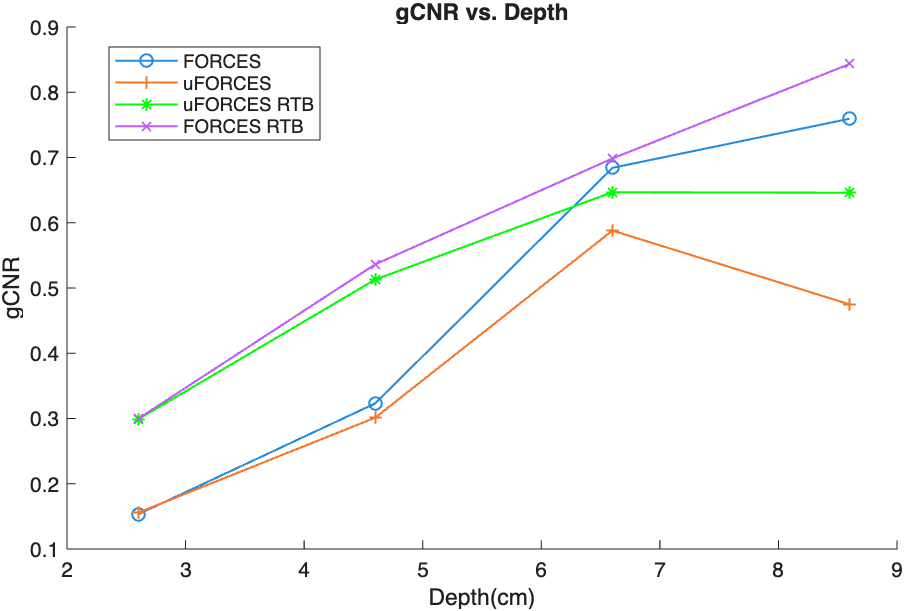}}
\caption{gCNR calculated for four cysts at various depths. The elevational focus is located 8.5 cm}
\label{fig6}
\end{figure}

Tubular anechoic cyst targets were imaged to compare the contrast in four different beamforming methods. Figure \ref{figcystref} shows the perpendicular view of the cysts analyzed in this study. Figure \ref{fig5} shows the four different images compared in this study. Generalized Contrast to Noise Ratio (gCNR) was used to quantify contrast performance of the different methods \cite{gCNR}. Figure \ref{fig6} shows the gCNR of FORCES, uFORCES, FORCES RTB, and uFORCES RTB plotted against the distance from the focal zone. Similar to the FWHM comparison, all methods perform well near the focal zone. However, both RTB methods perform better as the cysts move further from the focal zone. This improvement is most significant for the two cysts closest to the array with gCNR values approximately doubling when using RTB.

\section{DISCUSSION AND CONCLUSION}
\label{sec:discussionandconclusion}
This work demonstrates the improved elevational focusing achieved by utilizing FORCES RTB methods. When imaging a wire phantom, uFORCES RTB achieved better elevational resolution when compared to FORCES at nearly all distances from the focal zone. Although a uFORCES RTB image requires double the transmits of single FORCES image, one uFORCES RTB acquisition allows for the beamforming of 16 imaging planes, whereas a FORCES acquisition allows for the beamforming of only one plane. Notably, as we move further from the focal zone, the uFORCES RTB image experiences significant grating lobes. Given the virtual source spacing of 500 $\mu$m, the center frequency of 4.3 MHz, and the speed of sound of 1452 m/s, the effective pitch in the elevation direction is 1.5$\lambda$. Therefore, grating lobes are expected \cite{gratinglobes}. By reducing the spacing of the FORCES image planes acquired, the grating lobes could be reduced. Though, this would require either a reduction in resolution or a decrease in frame rate. Future work should explore these trade-offs.

When imaging the cyst phantom seen in Figure \ref{fig5}, FORCES RTB outperforms FORCES at all locations, and uFORCES RTB either outperforms or performs similarly to FORCES at all locations except for at the focal zone. While FORCES RTB requires significantly more transmits than FORCES, 16$\times$ the transmits in our study, we often utilize walking FORCES to create high resolution volumes when a high frame rate is not necessary \cite{FORCESvsRCA_PrePrint}. FORCES RTB should be utilized instead in these cases as it would require no additional data acquisition. 

Other work has demonstrated the value of volumetric ultrasound imaging for carotid stenosis detection and care \cite{CarotidProbe} \cite{PlackDetection3D}. Because RTB sequences in this study can beamform an 8 mm slab per transmit, the method could be well-suited for volumetric tasks such as carotid stenosis imaging, where volumetric coverage is critical. Future work should test this directly.

This work further demonstrates the flexibility of TOBE arrays. Other forthcoming methods such as HERCULES, which allows for receive focusing everywhere in a volume, can also be implemented on TOBE arrays \cite{dahunsi2025hadamardencodedrowcolumn}. This also includes other conventional RCA imaging methods such as Tilted Plane Wave and Virtual Line Source imaging \cite{FORCESvsRCA_PrePrint}. Beamforming was also performed offline in MATLAB for the purposes of this study. Due to the large number of transmits required, our current approach may be susceptible to motion artifacts. Future work is expected to address this issue. With the appropriate adjustments to our current system, uFORCES RTB can likely be implemented for real time imaging, along with the methods listed above. Future work should continue to implement these methods into a cohesive system. 

\section*{CONFLICTS OF INTEREST}
\label{sec:Conflicts}
RJZ and MRS are directors and shareholders of CliniSonix Inc., which provided partial support for this work. RJZ is a founder and director of OptoBiomeDx Inc., which, however, did not support this work. RJZ is also a founder and shareholder of IllumiSonics Inc., which, however, did not support this work.

\bibliographystyle{IEEEtran}
\bibliography{references}

\end{document}